# Is General Relativity a (partial) Return of Aristotelian Physics?

Herbert Pietschmann

Faculty of Physics, Univ. of Vienna.

Abstract:

Aristotle has split physics at the sphere of the moon; above this sphere there is no change except eternal spherical motion, below are two different kinds of motion: Natural motion (without specific cause) and enforced motion. In modern view motion is caused by gravity and by other forces. The split at the sphere of the moon has been definitely overcome through the observation of a supernova and several comets by Tycho Brahe. The second distinction was eradicated by Isaak Newton who showed that gravitational motion was caused by a force proportional to the inverse square of the distance. By the theory of General Relativity, Albert Einstein showed that there is no gravitational force but motion under gravity (i.e. Aristotle's "natural motion") is caused by the curved geometry of spacetime. In this way, the Aristotelian distinction between natural motion and enforced motion has come back in the form of two great theories: General Relativity and Quantum Field Theory which are today incompatible. To find a way out of this dilemma is the challenge of modern physics.

## 1. Aristotelian Physics

Aristotle – pupil of Plato – turned away from the philosophy of his teacher with the words:[1] "*And since Sokrates exerted himself about ethical matters and not at all about the whole of nature, but in the former sought the universal and was the first to be skilled at thinking about definitions, Plato, when he adopted this, took it up as applying to other things and not to sensible ones, because of this: it was impossible that there be any common definition of any of the perceptible things since they were always changing.*"

Consequently, Aristotle turned to "the whole of nature" and wrote the first comprehensive book about "Physics". A word of warning: The word "physics" has dramatically changed its meaning through the work of Galilei, Descartes, Newton (and others) in the 17th century. With Aristotle, "physics" means the philosophical description of "the whole of nature" not including humankind, for example ethics, aesthetics and the like. (It is not yet an "experimental science" which it only became in the 17th century.) Likewise, the word "Mathematics" has to be used carefully, since at the times of Plato and Aristotle, it was essentially geometry and number theory. Even in the first half of the 17th century, differential calculus and analytic geometry was not yet known.

Before we go into details (only as much as necessary for our purpose), let me quote Markus Fierz[2] who wrote a book on the history of mechanics. He says[3] about Aristotelian physics, "*it is a complete theory not without inner logics. In it, each phenomenon gets its plausible explanation thus forming an edifying picture of a well ordered cosmos. ... The theory is too empirical because it aims at a systematic representation of daily observations and experiences.*"

Among the many interesting details of Aristotelian physics two aspects are important for our discussion: First, Aristotle divided the cosmos into two distinct parts at the sphere of the moon. Below this sphere, everything was constituted by the four elements[4] air-fire-water-earth, above this sphere only the fifth element – quintessentia or ether – was present. Above the sphere of the moon, there was no change except circular movement and everything was eternal. Only movement in circles was possible.[5] "*We can now go on to claim that it is*

---

1 Aristotle: Metaphysics 987b. (Quoted from the translation of Joe Sachs.)

2 Successor of Wolfgang Pauli as professor of theoretical physics at ETH Zurich from 1960-1977.

3 M. Fierz: Vorlesungen zur Entwicklungsgeschichte der Mechanik. Springer Verlag Berlin (1972) p.13. German original: dass es sich um eine geschlossene, nicht ohne innere Logik aufgebaute Theorie handelt. In ihr findet jede Erscheinung eine einleuchtende Erklärung, wobei ein erbauliches Bild des wohlgeordneten Kosmos entsteht. ... Die Theorie ist allzu empirisch, denn sie möchte das, was man täglich beobachtet und erlebt ... systematisch darstellen.

4 It might be better to use the word „principle" instead of „element" because the latter has changed its meaning dramatically by modern chemistry and the periodic table.

5 Aristotle: Physics 261b27 & 265a. (Quoted from the translation of Robin Waterfield.)

*possible for there to be an infinite change, which is single and continuous, and that this is circular movement….our argument has now reached the point of making the general claim about every kind of change that none of them except circular movement can be continuous;…"*

Below the sphere of the moon, all kinds of change and motion were possible. But – and this is the second aspect to be considered – motions were divided into two categories: Natural motion ("motion moved by itself", without external force) and enforced motion due to an application of an external force. Aristotle:[6] *"Everything in motion is moved either by itself or by something else. Now, where self-movers are concerned it is obvious that the moved object and the agent of movement are contiguous; after all, the immediate agent is within the thing moved, so there is nothing in between. As for things that are moved by something else, there are four kinds of movement which are imparted by an external agent – pulling, pushing, carrying and rotating."*

An apple falling from a tree was a natural motion, a stone thrown into the air was enforced motion as long as it was in the hand of the thrower; letting loose was the transition to natural motion. However, Aristotle did not have the concept of conservation of energy and momentum. Therefore he had to make ad hoc assumptions to explain why the stone continued its motion in the direction of the hand of the thrower:[7] *"Given that, with the exception of self-movers, every moving object is moved by something, how is it that some things – things that are thrown, for instance – have continuity of movement when that which initiated the movement is no longer in contact with them?"*. (These ad hoc assumptions are of no importance for our considerations. The first vague idea of continuous energy came from Hipparchos of Nicaea, only about 190-120 BC.[8])

For enforced motion, Aristotle even had the law governing the connection between the applied force $F$ and the resulting velocity $v$.[9] Since he did not (could not) abstract from friction $W$, his law was

$$F = const.W.v \qquad\qquad (1)$$

Force is proportional to velocity which is also true in modern physics when we include friction and neglect acceleration (which at the time of Aristotle was never very important in daily life).

2. The first blow to Aristotelian physics by Tycho Brahe.

In 1572, a supernova occurred in our galaxy. Tycho Brahe was the leading astronomer at that time. Although he did not yet possess telescopes, his measurements were as precise as possible at the time. A "nova" (a new star), as Tycho Brahe coined the term, was impossible in the Aristotelian picture of the world, since beyond the sphere of the moon there was no change at all. John R. Christianson writes:[10] "*A great new star, a supernova, burst forth in the constellation Cassiopeia in 1572. Tycho observed it from Herrevad Abbey. His observations are famous in the history of science because he was able to prove for the first time, on the basis of empirical evidence, that change could and did occur in the celestial regions, just as on earth. This disproved a fundamental axiom of the Aristotelian world view, celestial immunability, and became an important first step in resolving the post Copernican crisis. In a short Latin treatise on the new star, Tycho stated quite unequivocally that "this new star is not located in the upper regions of the air just under the lunar orb, nor in any place closer to earth ... but far above the sphere of the moon in the very heavens, ... in the eighth sphere [of the fixed stars], or not far from there, in the upper orbs of the three superior planets," and he explicitly emphasized the revolutionary cosmological implications of this discovery.*"

Only five years later, on November 13, 1577 a comet appeared. For Tycho Brahe it confirmed his objection against Aristotelian cosmology. Christianson comments:[11] "*His studies of this comet became another milestone in the history of science. First of all, Tycho was able to establish that the comet was located above the sphere of the moon, confirming his conclusion from the supernova of 1572 that changes could and did occur in the heavens, as well as on earth. Second, Tycho demonstrated that the path of the comet cut across the orbits of several planets, proving they were not borne on solid crystalline spheres. This disproved another supposed tenet of Aristotelian cosmology and strengthened the evidence for a more dynamic, Neoplatonic cosmology.*"

Two more comets appeared soon after the first and were studied by Tycho Brahe. Christianson writes:[12] "*By 1586, Tycho Brahe has formed an elaborate plan for publishing his life's great masterwork .... The work would begin with three volumes under the general title <u>On the Most Recent Phenomena of the AEtherical World</u> (De mundi aetherei recentioribus phaenomenis.) The first volume would treat the new star of 1572, the second the comet of 1577, and the third the later comets. The very title asserted that these phenomena occurred in the celestial regions and not in the regions below the moon: These three volumes*"

---

*were to be an overwhelming attack on Aristotelian cosmology, based on massive, irrefutable analysis of empirical evidence and intended to bury the old view once and for all.*"

### 3. Newtons Unification of Physics.

A little story is very often told as an anecdote without deeper meaning: The young Isaac Newton is said to rest under an apple tree and when he observes an apple falling down from its branch he suddenly has the enlightenment that the force pulling the apple and the force holding the planets in their orbits are identical![13]

Obviously, the identity of these two forces is ingredient in Newtons Law of Gravitation. But the upper story points to a much deeper insight: It is the final overthrow of Aristotles division of physics into the two realms above and below the sphere of the moon. Hence it marks the beginning of the new epoch of science with validity in the whole universe.

But Newton also removed Aristotles division of the motions into "natural motions" and "enforced motions". For Newton, also Aristotles "natural motions" became "enforced" by the gravitational force. In this way, he unified our understanding of dynamics in the most general way. Later on it turned out, that the two forces $F$ governing our daily life, gravitation and electricity, were representable in a very similar way as gradient of a potential $V$ which is inverse proportional to the distance $r$.

$$F = - \text{grad } V \qquad\qquad (2a)$$

$$V = \text{Const.} (1/r) \qquad\qquad (2b)$$

Only the Constant is different for gravitational forces (Const. $= G.m_1.m_2$) and electrostatic forces (Const. $= Q_1.Q_2$) where $G$ is Newtons Gravitational constant, $m_i$ are the masses and $Q_i$ the electric charges of the participating bodies. Unlike Aristotles equation (1), Newtons "axiom" equates force with acceleration

$$F = m.a \qquad\qquad (3)$$

where $a$ is the acceleration of the mass $m$.

---

13 This story was first published 1752 by William Stukeley, one of Newtons first biographers in the form: "*After dinner, the weather being warm, we went into the garden and drank thea, under the shade of some apple trees...he told me, he was just in the same situation, as when formerly, the notion of gravitation came into his mind. It was occasion'd by the fall of an apple, as he sat in contemplative mood. Why should that apple always descend perpendicularly to the ground, thought he to himself...*" (royalsociety.org/turning-the-pages)
In another version Newton is said he was eager to tell this story because he was afraid that Hook would try to steel the copyright for the gravitational equation.

In the 20th century Quantum Mechanics was developed. In conjunction with special relativity it was extended to Quantum Field Theory which describes all forces relevant in the realm of particle physics, i.e. electromagnetic, weak and strong forces. Gravitational forces are too weak to be considered in particle physics. Thus the so called "Standard Model" of particle physics was developed and showed tremendous success when confronted with experiments.[14] In a sense, it was able to unify all three forces relevant in particle physics and when the famous "Higgs-boson" was discovered in 2012, our knowledge on the forces reigning among elementary particles seemed – at least temporarily – complete.

But gravitation did not take part in this game!

4. Einsteins Theory of Gravitation.

In 1915, Albert Einstein had found the basic equation in his theory of general relativity. He published a complete version of his thoughts in 1916.[15] According to this theory, the gravitational interaction was not caused by a force but by a curvature of spacetime. In this basic publication Einstein writes: "*Carrying out the general theory of relativity must lead to a theory of gravitation: for one can generate a gravitational field by merely changing the coordinate system.*"[16]

The astronomer and philosopher Arthur Stanley Eddington had verified Einsteins theory by measuring the deflection of the star light passing nearby the sun. This could only be done on the occasion of a solar eclipse in 1919. By this observation, he had also ruled out Newtons idea of a gravitational force albeit in a sector very far from daily experience.

In May 22, 1922, Eddington gave a very eloquent talk on Einsteins theory, entitled "The Theory of Relativity and its Influence on Scientific Thought".[17] In this talk he said: "*Einstein, recognizing that in the phenomena of gravitation he was not dealing with a 'tug' but with a curvature of the world, had to reconsider the law of gravitation. He could not make any possible law of curvature correspond exactly with the previously assumed law of tugging. Thus he was led to propound a new law of gravitation.*" Eddington also referred to his ruling out the Newtonian prediction with the words: "*I might ... remind you that ... the*

---

14 E.g. Dieter Haidt and Herbert Pietschmann: Electroweak Interactions - Experimental Facts and Theoretical Foundation. Landolt-Börnstein New Series Group I Vol. 10 (Springer Verlag, Berlin-Heidelberg-New York, 1988)

15 Albert Einstein: Die Grundlagen der allgemeinen Relativitätstheorie.
    Annalen d. Physik Serie6,**49**(1916)769-822.

16 German original: Die Durchführung der allgemeinen Relativitätstheorie muss zugleich zu einer Theorie der Gravitation führen: denn man kann ein Gravitationsfeld durch bloße Änderung des Koordinatensystems „erzeugen".

17 Arthur S. Eddington: Selected Works on the Implications of Relativity. Minkowski Inst. Press (2015)

*point at issue between Newton's and Einstein's theory was not the existence of a deflexion but the amount of the deflexion. Einstein predicting twice the deflexion possible on the Newtonian theory. The larger deflexion was quantitatively confirmed by the eclipse observations.*"

For Einstein, it was especially important to insist that there is no gravitational "force" (or "tug" in Eddingtons formulation) for such a force was not in any way observable. Already for the formulation of his special theory of relativity (1905) he rested his arguments on the fact that any "absolute time" was unobservable. In his paper on General Relativity from 1916 he writes:[18] "*the law of causality makes a sensible statement on the empirical world only when cause and effect are observable.*" Since a gravitational "force" was not observable, Einstein had eliminated it from his theory of gravitation and replaced it by the curvature of spacetime.

In this connexion it is historically interesting that only ten years later Einstein converted from these ideas which had led him to his most fundamental contributions to physics. Werner Heisenberg who had used the same philosophy for the derivation of his uncertainty relation, recalls a conversation with Einstein in 1926;[19] Einstein: "*You don't seriously believe that a physical theory can only contain observables.*" Heisenberg: "*I thought you were the one who made this idea the foundation of your theory of relativity?*" Einstein: "*May be I have used this kind of philosophy, nevertheless it is nonsense.*"

At this point we might reflect a little bit on the general state of mind in our society. Clearly, the general public (including many physicists) still "believes" in the existence of the gravitational force, although it has definitely been ruled out by Eddingtons experiment (and many more precise experiments since then). Since the Theory of General Relativity is mathematically and intellectually extremely challenging, this may be understandable and acceptable. More so, because practically every situation we meet in daily life can also be interpreted with Newtons gravitational force without observable deviations. But there are counterexamples! Any satellite navigation system (e.g. GPS – global positioning system) would mislead its user if it were based on Newton's gravitational force; Einsteins General Relativity, i.e. the curvature of spacetime  is required as basis for the prediction of positions.[20]

18 German original: … das Kausalitätsgesetz hat nur dann den Sinn einer Aussage über die Erfahrungswelt, wenn als Ursachen und Wirkungen letzten Endes nur *beobachtbare Tatsachen* auftreten.

19 W. Heisenberg: Der Teil und das Ganze. Piper Verlag München (1969) 91f.
German original: A.E.: Sie glauben doch nicht im Ernst, dass man in eine physikalische Theorie nur beobachtbare Größen aufnehmen kann.
W.H.: Ich dachte, dass gerade Sie diesen Gedanken zur Grundlage Ihrer Relativitätstheorie gemacht hätten?
A.E.: Vielleicht habe ich diese Art von Philosophie benützt, aber sie ist trotzdem Unsinn.

20 The situation is somewhat similar to atomic physics. Most schools teach in their classes that the atom is kind of a tiny planetary system with the nucleus in the center and electrons moving around it like tiny planets. This is still widely accepted knowledge though we know since 1926 (the foundation of Quantum Mechanics) that it is To summarize, let us state that motion caused by gravitation is not caused by a force; in that sense it differs from all other motions. Einstein made this clear in his quoted paper from 1916. He writes:[21] "*According to the theory of General Relativity gravitation has an exceptional role with respect to all other forces, especially electromagnetism.*" This is what in a sense we might call "return of Aristotelian physics" since it clearly distinguishes between "natural motion" and "enforced motion", constituting the basic problem of modern physics. Acceleration is either caused by the geometry of spacetime (gravitation) or by an external force in Euclidian spacetime (all other forces). Mathematically, these two different views are represented either by the Theory of General Relativity (gravitation) or by Quantum Field Theory (all other forces).[22]

## 5. The Challenge of Modern Physics

In present day physics we have two different theories (General Relativity and Quantum Field Theory) who are both extremely well tested and corroborated by a multitude of precise experiments. But they are incompatible! Quantization of General Relativity has so far failed in all attempts.[23] The situation is well described by Joseph Polchinski in a paper called "Burning Rings of Fire".[24] He refers to General Relativity and Quantum Field Theory and writes: "*It seems that physicists must give up one of our widely cherished beliefs, but we cannot agree on which one. We hope, however, that out of this confusion will come ... a way to finally resolve the apparent contradiction between these two reigning theories of physics.*"[25]

Qualitatively, we can compare the situation with Aristotelian physics: There is "natural motion", motion under gravity, which is the straightest possible motion in a curved spacetime; and there is "enforced motion" due to an application of

_______________________

totally wrong and helplessly false!

21 German original: Die Gravitation spielt also gemäß der allgemeinen Relativitätstheorie eine Ausnahmerolle gegenüber den übrigen, insbesondere den elektromagnetischen Kräften …

22 In Quantum Field Theory, an accelerated charged particle radiates. According to the elevator gedankenexperiment of Einstein, a charged particle accelerated purely by gravitation should not radiate. This contradiction could be cleared by experiment.

23 String theory is sometimes quoted to be the solution to this problem. However, string theory is not a physical theory in the sense that it cannot predict a test experiment which could be performed. Hence it is applied mathematics.

24 Sci. Am. April 2015 Vol.312/4 p.23-27.

25 The apparent contradiction does not prevent computations in areas where both theories are relevant. However, these computations have to be done on the basis of "effective theories" rather than the full theories of General Relativity and Quantum Field Theory. See e.g. John F. Donoghue: The effective field theory treatment of quantum gravity. arXiv:1209.3511v1 [gr-qc] 16 Sept. 2012.

an external force.[26] (Obviously, the mathematical description of both kinds of motion are far beyond anything possible at the time of Aristotle.)

What we need is a unification of the nature of Newton's scientific revolution. But the great power of todays theories makes it even more difficult to achieve that goal. Therefore we have to approach it step by step. The first step is the empirical establishment of gravitational waves which is under way and should produce an answer soon.[27] The next step would probably be to answer the question whether particles[28] associated with gravitational waves[29] have experimentally accessible consequences. (If they are experimentally found). We can only hope that this will eventually lead to ideas that will give birth to a new theory comprising both Gravitation and Quantum Field Theory, in this way describing all kinds of motion in a unified manner.

The author thanks Franz Embacher for reading the manuscript.

---

26 In technical terms:Interaction with a virtual particle.

27 On Feb. 11, 2016, a press conference was held by the LIGO-collaboration to announce the observation of a gravitational wave. In order to speak of a discovery, we have to wait for at least one reproduction (see e.g. Herbert Pietschmann: Phänomenologie der Naturwissenschaft. Ibera/European University Press, Vienna (2007) Chap. 6/6.)

28 so called gravitons

29 required by Quantum Field Theory